# Harmonics of the AC susceptibility for the study of I-V curves in melt grown YBCO

*Maria Giuseppina Adesso, Massimiliano Polichetti, Sandro Pace*

Dipartimento di Fisica – INFM Universita' degli Studi di Salerno, Via S. Allende, 84081 Baronissi (Salerno), ITALY

**Abstract**

The measurements of the 1$^{st}$ and the 3$^{rd}$ harmonics of the AC magnetic susceptibility, performed on melt grown YBCO samples as a function of the temperature and the external parameters (frequency and amplitude of AC magnetic field, intensity of a DC field) were studied by using the Cole-Cole plot analysis. By considering the AC field dependence of the 3$^{rd}$ harmonics Cole-Cole plots, we have extracted information about the peculiar behaviour of the I-V characteristics and, therefore, about the dissipative phenomena in the sample. The experimental results have been analysed with the help of both analytical and numerical curves. In particular, the use of a phenomenological expression for the resistivity allowed us to reproduce the main experimental features, including the observed invariance of the area inside the Cole-Cole plots and their behaviour as a function of the AC field.

**Introduction**

The non-linear electro-dynamical behaviour of the superconductors has been widely studied by two complementary approaches: the magnetic techniques and the transport measurements. Specifically, both the AC magnetic susceptibility [1,2] (and in particular its higher harmonics) and the Current-Voltage (I-V) curves [3] allow us to obtain information about the pinning potential and the different dissipative phenomena governing the magnetic/electrical sample response.

Nevertheless, the magnetic susceptibility is an indirect experimental technique, and therefore, in order to interpret the measurements, we need to compare them with analytically and/or numerically computed curves, where assumptions about the dissipative regimes and the pinning model have to be done [4]. By comparing the experimental data with those evaluated numerically [4], it has been shown that, for a given vortex dynamics, the temperature dependence of the 3rd harmonics is very sensitive to the different temperature dependencies of the pinning potential [5]. Moreover, if the dominant pinning has been individuated, the behaviour of the $3^{rd}$ harmonics can be interpreted in terms of the different types of dissipative phenomena (Flux Creep, Flux Flow, Thermally Activated Flux Flow) [6].

On the contrary, the measurements of the Current-Voltage characteristics allow us to directly detect the specific type of dissipative regime but, in any case, in order to get information about the pinning potential, it is necessary to use some models which include the dependencies of the pinning potential on the current, on the temperature and on an eventually applied magnetic field [7].

In this work, we interpret our AC susceptibility measurements with the help of numerical simulations. The numerical data are obtained by choosing a non linear phenomenological resistivity, analogous to that used often to fit the experimental I-V curves [7], in which the temperature dependence of the pinning potential is deduced from the analysis of the harmonics of the AC susceptibility [5,6].

A key point of the present work is the introduction of the analysis of the 3$^{rd}$ harmonics Cole-Cole plots as an instrument to probe the critical choice of the phenomenological parameters, and in particular the *n*-value, in the models describing the I-V characteristics.

**Experimental results**

We measured the temperature dependence of the first ($\chi_1^{''}$) and the third ($\chi_3^{''}$) harmonics of the AC susceptibility, by an home made AC susceptometer, on some melt grown $Y_1Ba_2Cu_3O_{7-\delta}$ samples, at different amplitudes and frequencies of the AC magnetic field, with or without an external DC magnetic field, both fields being parallel to the longitudinal axis of the sample.

In this article, we mainly refer to the measurements of the 3$^{rd}$ harmonics, performed at different amplitudes of the AC field ($h_{AC}$ = 4, 8, 16, 32 Oe), in absence of a DC field and at a fixed frequency ($\nu$ = 107 Hz), on an homogeneous slab with dimensions 1.8 mm x 3.2 mm x 5 mm.

In Fig.1 the Cole-Cole plots of the measured 3rd harmonics are shown, and the area S inside the curves is also reported for each amplitude. In the inset of Fig.1, the temperature dependencies of the real part of the 3$^{rd}$ harmonics, at the same external conditions, are plotted.

From these data, we can observe that the shape of the Cole-Cole plots of the 3rd harmonics is quite different if $h_{AC}$ is changed, as it happens for the shape of each individual components as function of the temperature. In particular, if $h_{AC}$ increases, the polar plots turn in the direction of the left semi-plane, corresponding to more and more negative values of $\chi_3'(T)$. In spite of these evident changes, the area of the Cole-Cole plots is constant, within the 1% experimental errors, namely S=(1.01±0.01)x10$^{-3}$.

In order to interpret these behaviours in terms of the hysteretic phenomena and/or of the non-linear dynamic regimes, we computed the 3rd harmonics of the AC susceptibility both analytically and numerically.

**Analytical results**

In Fig.2, the temperature dependence of the real part (on the left) and the Cole-Cole plots (on the right) of the 3rd harmonics, analytically computed by using the Bean model [8], are plotted for different amplitudes of the AC magnetic field.

As we can observe, in the Bean model the Cole-Cole plots occupy only the right semi-plane, corresponding to only non-negative $\chi_3'(T)$ values, whereas the experimental $\chi_3'(T)$ are characterized by both negative and positive values. Nevertheless, by increasing the amplitude of the AC field, neither the shape nor the area (S=2.65x10$^{-3}$) of the Cole-Cole plots change, although the corresponding $\chi_3'(T)$ peaks curves shift towards the lower temperatures.

**Numerical results**

The comparison between the experimental and the analytical results suggests that, in order to interpret the shape variations of the 3$^{rd}$ harmonics, it is necessary to consider the vortex dynamics effects. Therefore, by using a technique already used in previous works [4-6] we integrated the non-linear diffusion equation of the magnetic field [4], in the present case choosing a phenomenological resistivity, obtained by supposing that the pinning potential, $U_p(J,T,h_{AC})$, depends on the current through a logarithmical law, and on the $h_{AC}$ through a power law:

$$U_p(J,T,h_{AC}) = A \cdot g(T) \cdot h_{AC}^{-\beta} \cdot \ln(J_c(T)/J) . \qquad (1)$$

Here A is some proportionality constant, as suggested by Zeldov [7] who fitted experimental I-V curves with the help of similar assumptions where the dependence on the intensity of a DC field (absent in our analysed measurements) is considered instead of the amplitude of the AC magnetic field.

The temperature dependence of $g$ and $J_c$ are chosen according to the δl type collective pinning model, as suggested in the References [5,6,9]:

$$g(T) = 1 - t^4 \qquad (2)$$

$$J_c(T) = J_c(0)(1-t^2)^{5/2} \cdot (1+t^2)^{-1/2} \qquad (3)$$

where $t = T/T_c$.

With the help of (1), the resistivity can be written as:

$$\rho = \rho_0 \cdot e^{-\frac{U_p(J,h_{AC},T)}{k_B T}} = \rho_0 \cdot (J/J_c(T))^{n(T,h_{AC})} \qquad (4)$$

where $\rho_0$ is the Flux Flow resistivity [10] and $n(T,h_{AC})$ is a coefficient which depends on $T$ by means of the pinning model:

$$n(T,h_{AC}) = A \cdot \frac{g(T)}{T} \cdot h_{AC}^{-\beta} \qquad (5)$$

The values of the parameters used in the simulations are: $A = 1.6 \times 10^4$ K/Oe, $T_c = 91.6$ K, and $J_c(0) = 10^9$ A·m, as experimentally estimated for our YBCO samples [6].

In Fig. 3), 4) and 5), the simulated Cole-Cole plots are reported for different $\beta$-values. From these figures, it is possible to deduce that the choice of the $\beta$ exponent is critical, and this is particularly evident if we observe the $h_{AC}$ dependence of the 3$^{rd}$ harmonics Cole-Cole plots; in particular, if we change the $\beta$ values, we can identify three completely different behaviours, summarized as follows:

*a)* For $\beta < 0.55$ (in Fig.3 the curves for $\beta = 0.25$ are plotted, but similar results have been obtained with other values in the same interval), the shapes of the 3$^{rd}$ harmonics Cole-Cole plots for increasing $h_{AC}$ change, but they turn towards the right semi-plane, contrary to what happens for the experimental curves, and their area grows.

*b)* For $\beta = 0.55$ (Fig.4), corresponding to a situation which is similar to the Bean critical state, by varying $h_{AC}$ neither the shapes nor the area (S=(1.45±0.02)x10$^{-3}$) of the Cole-Cole plots change significantly (about 1%), although each individual component of the 3$^{rd}$ harmonics (not reported) show more evident modifications. Nevertheless, contrary to the Bean critical state, the Cole-Cole plots lay in both the left and the right semi-plane.

*c)* For $\beta > 0.55$ (see Fig.5 for $\beta = 0.65$, although we obtained similar results by using other values), the Cole-Cole plots move towards the left semi-plane when $h_{AC}$ increases, according to the experimental data. Moreover, the area is constant within the 2% and, in particular, the value S=(1.28±0.03)x10$^{-3}$, closer to the experimental value, corresponding to $\beta = 0.65$, suggests that a further optimisation of the $\beta$ value will make the numerical data comparable with the experimental ones, permitting a better evaluation of the *n* parameter, which is fundamental in several power applications. Nevertheless, from this analysis, we can state that also small variations in the values of the $\beta$ parameter in the I-V curves allow to discriminate univocally among magnetic responses which behave in very different ways. We can suppose that the individuated behaviours could be related to different types of the flux dynamics, and a further effort is needed to analyze each individual dissipative regime.

**Conclusions**

In this work, we analysed the measurements of the 3$^{rd}$ harmonics of the AC susceptibility, performed on YBCO melt-grown samples, both analytically and numerically. We showed that more information can be obtained if the Cole-Cole plots of the 3$^{rd}$ harmonics are considered, instead of the temperature dependence of each individual 3$^{rd}$ harmonic components. In our measurements, the Cole-Cole plots as a function of the AC amplitude of the magnetic field $h_{AC}$ are characterized by shape variations with a tendency to occupy the left semi-plane when $h_{AC}$ increases, but by preserving a nearly constant area. Although the simple, analytical, Bean model seems to reproduce the area invariance, it is necessary to use numerical techniques in order to reproduce the experimental shape variations. The numerical curves were obtained by using a resistivity, deduced from the Current-Voltage characteristics models, in which the $h_{AC}$ dependence is included as a power law of a phenomenological parameter $\beta$. We showed that, for a given temperature dependence of the pinning potential, the choice of the $\beta$ parameter is critical and the experimental behaviour can be reproduced only if the right $\beta$ value is used in the numerical calculations. In this sense, the Cole-Cole plots represent a valid technique for determining the best values of the phenomenological parameters in the Current-Voltage characteristics.

**Acknowledgments**

The authors are very grateful to Dr. D. Zola and Dr. C. Senatore for the useful discussions, to Dr. A. Vecchione for the YBCO samples, and to Mr. A. Ferrentino for his technical support.

**Figures**

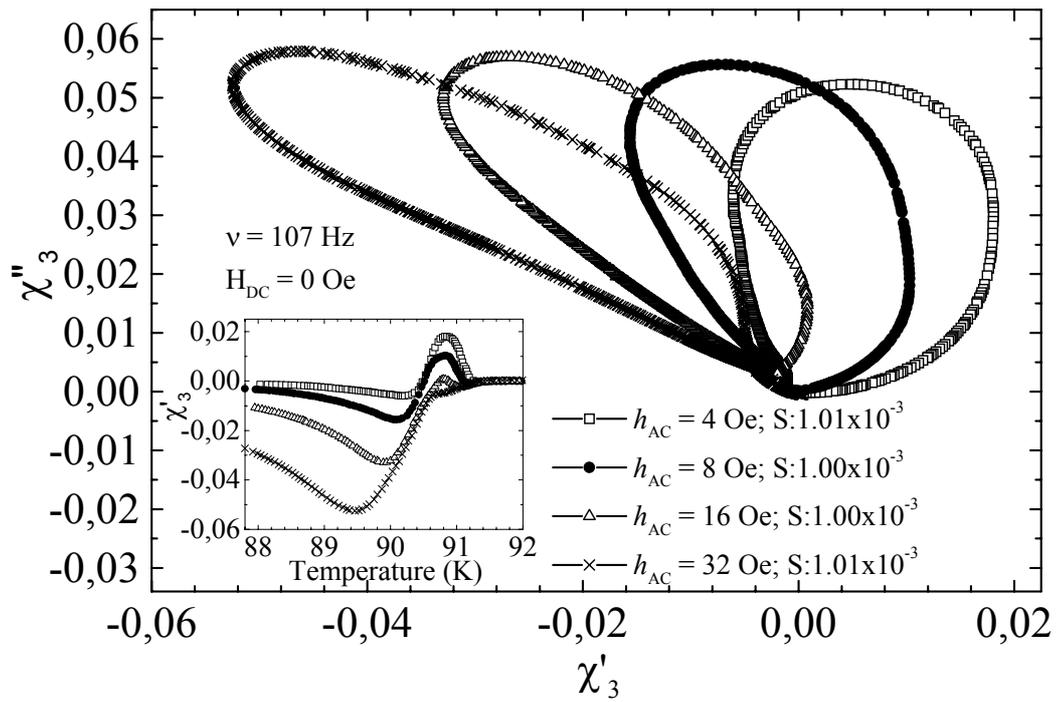

**Fig. 1** Experimental Cole-Cole plots of the 3$^{rd}$ harmonics at different $h_{AC}$ ($h_{AC}$= 4,8,16,32 Oe), at a fixed frequency $\nu$ = 107 Hz, without a DC field. The computed areas, S, are also reported for the different fields. The inset shows the temperature dependence of the real part of the 3$^{rd}$ harmonics at the same external conditions.

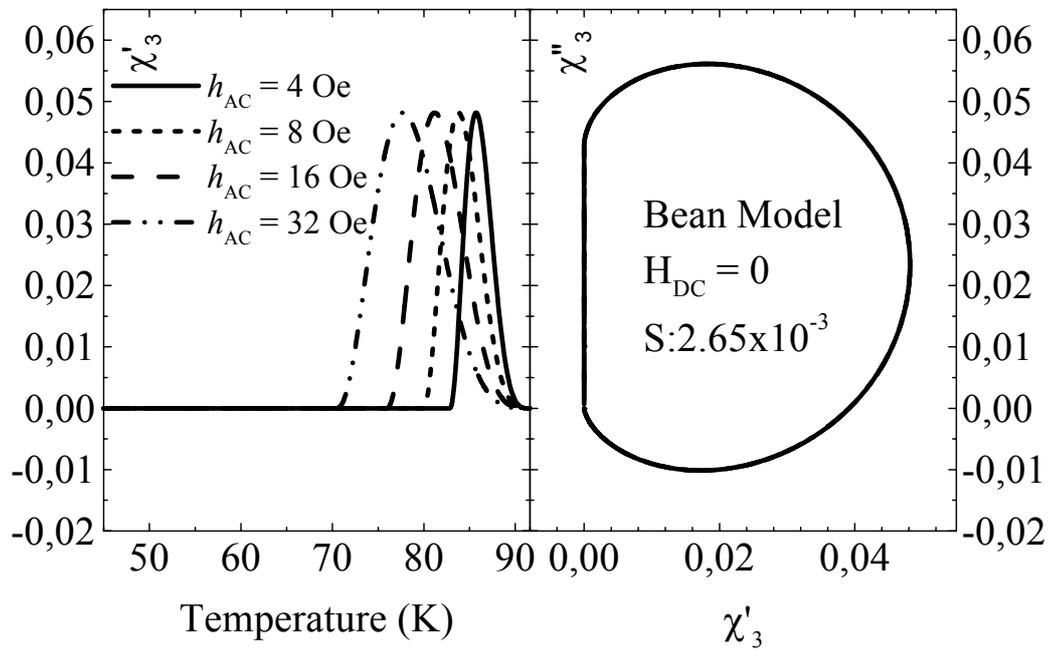

**Fig. 2** Temperature dependence of the real part (on the left) and Cole-Cole plots (on the right) of the 3$^{rd}$ harmonics, computed by using the Bean model: by increasing $h_{AC}$, the Cole-Cole plots are constant, whereas $\chi'_3(T)$ peak shifts to lower temperature

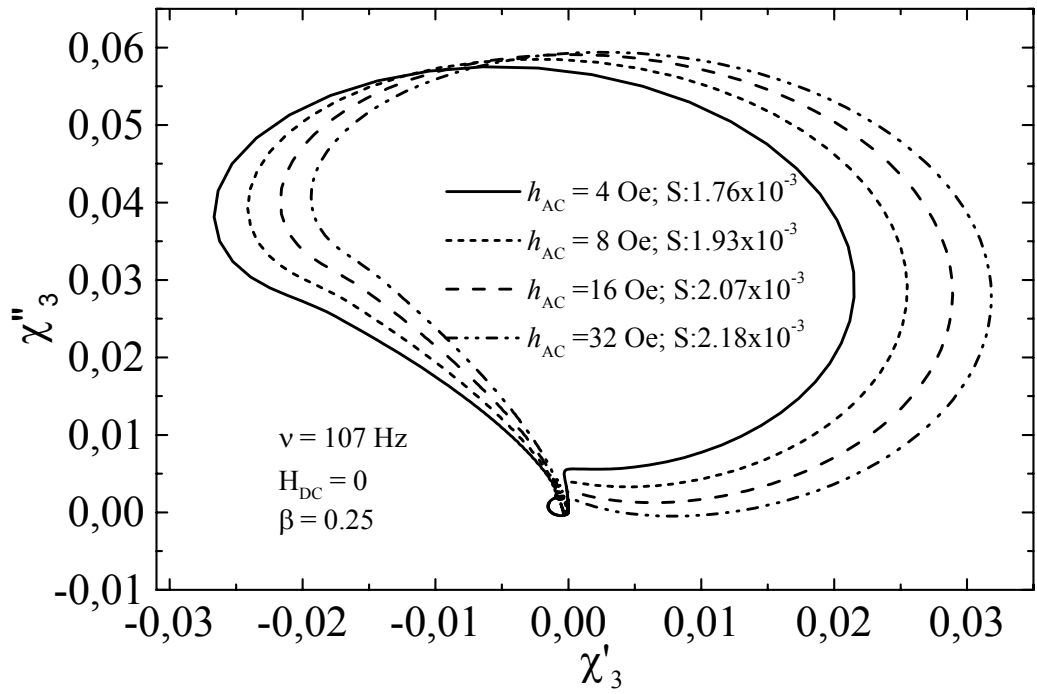

**Fig. 3** Cole-Cole plots of the 3$^{rd}$ harmonics computed numerically with $\beta$=0.25

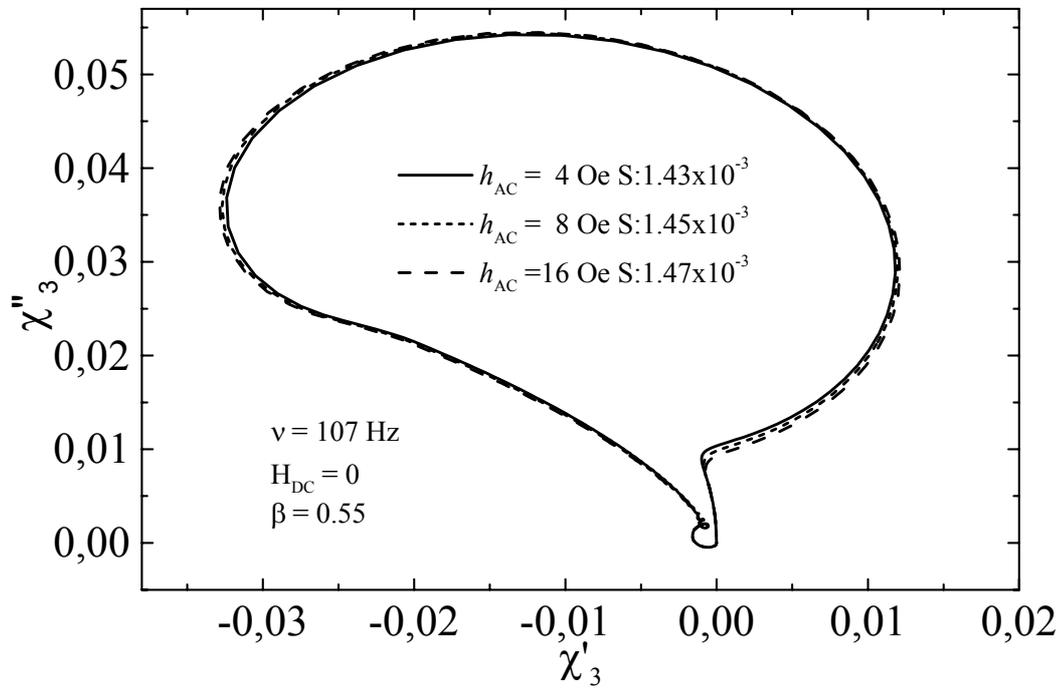

**Fig. 4** Cole-Cole plots of the 3$^{rd}$ harmonics computed numerically with $\beta$=0.55

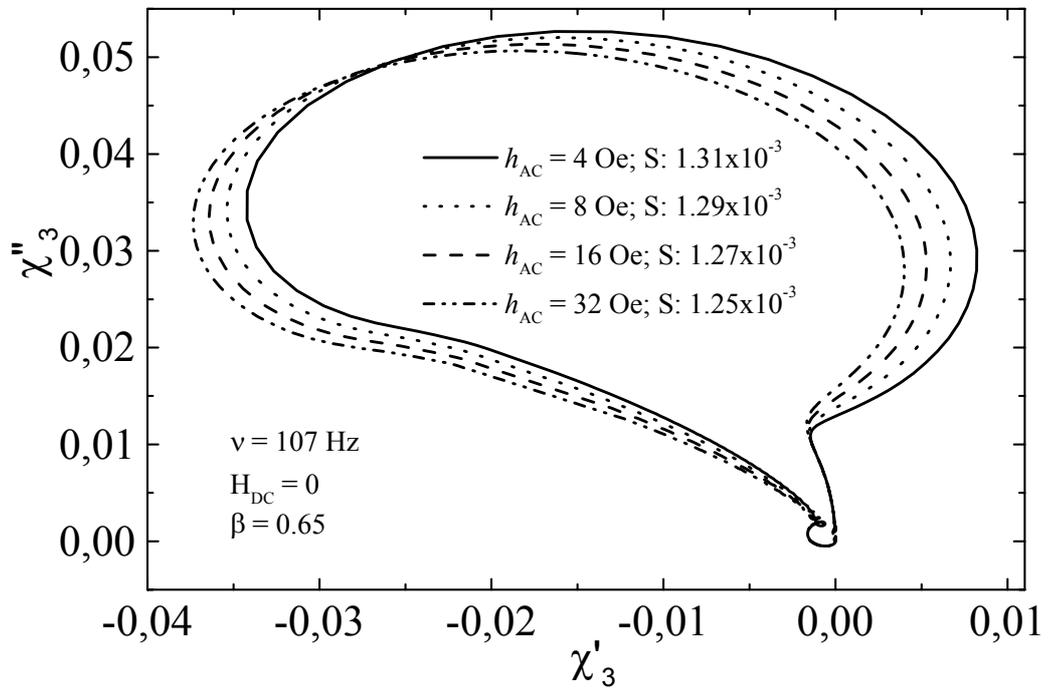

**Fig. 5** Cole-Cole plots of the 3$^{rd}$ harmonics computed numerically with $\beta$=0.65